\documentstyle[12pt,twoside,epsf]{article}

\def\oversim#1#2{\lower4pt\vbox{\baselineskip0pt \lineskip1.5pt
            \ialign{$\mathsurround=0pt#1\hfil##\hfil$\crcr#2\crcr\sim\crcr}}}
\begin{document}
\pagestyle{plain}
%

{\begin{center}
\Huge \bf                 Search for the Supersymmetric \\
			  Partner of the Top-Quark \\
          in $p \overline{p}$ Collisions at $\sqrt{s} = 1.8 \, {\rm TeV}$
\end{center} }
{
\font\eightit=cmti8
\def\r#1{\ignorespaces $^{#1}$}
\hfilneg
\begin{sloppypar}
\noindent
T.~Affolder,\r {23} H.~Akimoto,\r {45}
A.~Akopian,\r {38} M.~G.~Albrow,\r {11} P.~Amaral,\r 8 S.~R.~Amendolia,\r {34} 
D.~Amidei,\r {26} K.~Anikeev,\r {24} J.~Antos,\r 1 
G.~Apollinari,\r {11}
% ...
\end{sloppypar}
\vskip .026in
\begin{center}
(CDF Collaboration)
\end{center}

\vskip .026in
\begin{center}
\r 1  {\eightit Institute of Physics, Academia Sinica, Taipei, Taiwan 11529, 
Republic of China} \\
\r 2  {\eightit Argonne National Laboratory, Argonne, Illinois 60439} \\
\r 3  {\eightit Istituto Nazionale di Fisica Nucleare, University of Bologna,
I-40127 Bologna, Italy} \\
\r 4  {\eightit Brandeis University, Waltham, Massachusetts 02254} \\
\r 5  {\eightit University of California at Davis, Davis, California  95616} \\
\r 6  {\eightit University of California at Los Angeles, Los 
Angeles, California  90024} \\  
\r 7  {\eightit Instituto de Fisica de Cantabria, CSIC-University of Cantabria, 
39005 Santander, Spain} \\
\r 8  {\eightit Enrico Fermi Institute, University of Chicago, Chicago, 
Illinois 60637} \\
\r 9  {\eightit Joint Institute for Nuclear Research, RU-141980 Dubna, Russia}
\\
\r {10} {\eightit Duke University, Durham, North Carolina  27708} \\
\r {11} {\eightit Fermi National Accelerator Laboratory, Batavia, Illinois 
60510} \\
\r {12} {\eightit University of Florida, Gainesville, Florida  32611} \\
\r {13} {\eightit Laboratori Nazionali di Frascati, Istituto Nazionale di Fisica
               Nucleare, I-00044 Frascati, Italy} \\
\r {14} {\eightit University of Geneva, CH-1211 Geneva 4, Switzerland} \\
\r {15} {\eightit Glasgow University, Glasgow G12 8QQ, United Kingdom}\\
\r {16} {\eightit Harvard University, Cambridge, Massachusetts 02138} \\
\r {17} {\eightit Hiroshima University, Higashi-Hiroshima 724, Japan} \\
\r {18} {\eightit University of Illinois, Urbana, Illinois 61801} \\
\r {19} {\eightit The Johns Hopkins University, Baltimore, Maryland 21218} \\
\r {20} {\eightit Institut f\"{u}r Experimentelle Kernphysik, 
Universit\"{a}t Karlsruhe, 76128 Karlsruhe, Germany} \\
\r {21} {\eightit Center for High Energy Physics: Kyungpook National
University, Taegu 702-701; Seoul National University, Seoul 151-742; and
SungKyunKwan University, Suwon 440-746; Korea} \\
\r {22} {\eightit High Energy Accelerator Research Organization (KEK), Tsukuba, 
Ibaraki 305, Japan} \\
\r {23} {\eightit Ernest Orlando Lawrence Berkeley National Laboratory, 
Berkeley, California 94720} \\
\r {24} {\eightit Massachusetts Institute of Technology, Cambridge,
Massachusetts  02139} \\   
\r {25} {\eightit Institute of Particle Physics: McGill University, Montreal 
H3A 2T8; and University of Toronto, Toronto M5S 1A7; Canada} \\
\r {26} {\eightit University of Michigan, Ann Arbor, Michigan 48109} \\
\r {27} {\eightit Michigan State University, East Lansing, Michigan  48824} \\
\r {28} {\eightit University of New Mexico, Albuquerque, New Mexico 87131} \\
\r {29} {\eightit The Ohio State University, Columbus, Ohio  43210} \\
\r {30} {\eightit Osaka City University, Osaka 588, Japan} \\
\r {31} {\eightit University of Oxford, Oxford OX1 3RH, United Kingdom} \\
\r {32} {\eightit Universita di Padova, Istituto Nazionale di Fisica 
          Nucleare, Sezione di Padova, I-35131 Padova, Italy} \\
\r {33} {\eightit University of Pennsylvania, Philadelphia, 
        Pennsylvania 19104} \\   
\r {34} {\eightit Istituto Nazionale di Fisica Nucleare, University and Scuola
               Normale Superiore of Pisa, I-56100 Pisa, Italy} \\
\r {35} {\eightit University of Pittsburgh, Pittsburgh, Pennsylvania 15260} \\
\r {36} {\eightit Purdue University, West Lafayette, Indiana 47907} \\
\r {37} {\eightit University of Rochester, Rochester, New York 14627} \\
\r {38} {\eightit Rockefeller University, New York, New York 10021} \\
\r {39} {\eightit Rutgers University, Piscataway, New Jersey 08855} \\
\r {40} {\eightit Texas A\&M University, College Station, Texas 77843} \\
\r {41} {\eightit Texas Tech University, Lubbock, Texas 79409} \\
\r {42} {\eightit Istituto Nazionale di Fisica Nucleare, University of Trieste/
Udine, Italy} \\
\r {43} {\eightit University of Tsukuba, Tsukuba, Ibaraki 305, Japan} \\
\r {44} {\eightit Tufts University, Medford, Massachusetts 02155} \\
\r {45} {\eightit Waseda University, Tokyo 169, Japan} \\
\r {46} {\eightit University of Wisconsin, Madison, Wisconsin 53706} \\
\r {47} {\eightit Yale University, New Haven, Connecticut 06520} \\
\r {(\ast)} {\eightit Now at Carnegie Mellon University, Pittsburgh,
Pennsylvania  15213}
\end{center}
}
 
{\small
   We report on a search for the supersymmetric partner of the top quark
(stop) produced in $t \overline{t}$ events using $110 \, {\rm pb}^{-1}$
of $p \overline{p}$ collisions at $\sqrt{s} = 1.8 \, {\rm TeV}$ recorded
with the Collider Detector at Fermilab. In the case of a light stop
squark, the decay of the top quark into stop plus the lightest supersymmetric
particle (LSP) could have a significant branching ratio.
The observed events are consistent with Standard Model $t \overline{t}$
production and decay. Hence, we set limits on the branching ratio of the
top quark decaying into stop plus LSP, excluding branching ratios above 45\% 
for a LSP mass up to $40$ ${\rm GeV/c}^{2}$.

\pagebreak}

   With the observation in 1995 of a heavy top quark~\cite{cdf_top,d0_top}, 
an important prerequisite was met for low energy supersymmetry 
(SUSY)~\cite{general_susy} to explain electroweak
symmetry breaking. 
In the Minimal Supersymmetric extension of the
Standard Model (MSSM), all known 
particles of the Standard Model (SM) 
acquire supersymmetric partners, or superpartners. For 
fermions the superpartners are bosons, while for
bosons the superpartners are fermions. 
We assume conservation of a multiplicative
quantum number $\mathcal{R}$-parity,  
which requires these new particles to be produced in pairs and
prevents decays of the lightest supersymmetric particle (LSP). From
cosmological considerations~\cite{ellis_hagelin}, the LSP is normally 
assumed to be the lightest neutralino.


   The large\, Yukawa\, coupling\, of the top quark opens\, up the possibility
of a large mass 
splitting in the third generation of 
fermionic superpartners (the squarks and sleptons).
The superpartners of the right-handed and left-handed 
top quark (the stop squarks) combine to form the mass eigenstates.
The lightest stop squark ($\tilde{t}_{1}$) could then be lighter than the
superpartners of all other squarks. Most limits on squark masses
\cite{cdf_d0_squark} do not apply to the stop squark because they are
usually based on a model of five degenerate squarks. 
Current stop squark mass limits are significantly lower~\cite{lep_slc_stop}
than these limits or based on the assumption of 
a very heavy chargino ($\tilde{\chi}^{\pm}_{1}$)~\cite{d0_clsp}. 
The latter searches are complementary to the analysis presented here
since the stop decay mode  $c + \mbox{LSP}$ does
not coexist with the decay mode 
$b + \tilde{\chi}^{\pm}_{1}$.
   If the stop squark is light, decays of the top quark into stop plus
the lightest neutralino could be kinematically allowed. If this neutralino
is the LSP it will be stable and only weakly
interacting. Such a particle would pass through the detector without
interaction, causing a considerable energy imbalance. For the stop squark
we assume decays analogous to the Standard Model top quark decay,
i.e.\ into chargino and $b$-quark. The chargino could then decay into 
a LSP plus either a quark-antiquark pair or a lepton and neutrino.

   Branching ratios as large as $40$ to 50\% 
for the top decay into stop  have been suggested~\cite{stop_branching}. 
In such scenarios about one half of $t \overline{t}$ events would have 
one SM and one supersymmetric top decay. 
If the SM top decay to a W and a $b$-quark is
followed by the leptonic decay of the W, then                           
the selection criteria of the online leptonic
trigger, the offline dataset selection and the
$t \overline{t}$ event identification~\cite{cdf_top} will all be
satisfied. The decay of the second top quark 
can then be used to search for decays into stop.

   The CDF detector~\cite{cdf_1a_detector} is well suited to search for
supersymmetric top quark decays. The following components are relevant
to this analysis: the central tracking chamber, which is inside a 
$1.4 \,{\rm T}$ superconducting solenoidal magnet, measures the momentum of charged
particles with a resolution of 
$\delta p_{\rm T} / p_{\rm T} = 0.001 * p_{\rm T}$ ($p_{\rm T}$ in ${\rm GeV/c}$)~\cite{cdf_coord}. The silicon
vertex detector, with an inner radius of $3 \, {\rm cm}$ and an outer of
$8 \, {\rm cm}$, identifies secondary vertices with a resolution 
of $130 \, \mu {\rm m}$ in the transverse plane.
The electromagnetic and hadronic calorimeters cover the pseudorapidity
region $| \eta | < 4.2$ and are used to identify jets and electrons, and to
measure the missing transverse energy $\;\not\!\!\!E_{\rm T}$~\cite{cdf_coord}.
An outer layer of drift chambers provides muon identification in the region
$| \eta | < 1.0$.

   The search reported here is based on $110 \, {\rm pb}^{-1}$ of
$p \overline{p}$ collisions at $\sqrt{s} = 1.8 \, {\rm TeV}$ recorded
with the Collider Detector at Fermilab during the 1992-93 and 1994-95
collider periods. The analysis is a combination of a CDF single
lepton plus jet top analysis~\cite{cdf_top} and a kinematic
analysis~\cite{cdf_top_kin}.
The analysis cuts are slightly revised to improve sensitivity.

   The leptonic W decay from the SM top decay yields an energetic
lepton. Events are selected as in the single lepton plus $b$-jet
top analysis by requiring a central electron ($| \eta | \leq 1.1$) or
central muon ($| \eta | \leq 1.0$) with transverse momentum 
$p_{\rm T} \geq 20 \, {\rm GeV/c}$.

   The neutrino from the W decay, as well as any LSP's, will escape the
apparatus without detection, resulting in an energy imbalance. 
The  $\,\not\!\!\!E_{\rm T}$ measured by the
calorimeter is corrected if the lepton is a muon. We have increased the
$\,\not\!\!\!E_{\rm T}$ requirement, from $\,\not\!\!\!E_{\rm T} \geq 25 \, {\rm GeV}$ 
in the single lepton plus $b$-jet top analysis 
to $\not\!\!E_{\rm T} \geq 45 \, {\rm GeV}$, as our signal
is expected to have a harder $\not\!\!\!E_{\rm T}$ spectrum. 
To reject non-W background,
we require the transverse mass $M_{\rm T}$ of the lepton and $\,\not\!\!\!E_{\rm T}$ 
system to be larger than $40 \, {\rm GeV}$. 
While the top analyses are concerned with separating $t \overline{t}$ from W
plus jet production, in this analysis we are interested in 
minimizing W plus multijet background 
and then focusing on
separating SUSY top decays from SM top decays. In both SUSY and SM top
decays the W has a substantial transverse momentum. 
An additional requirement that the  lepton$-\not\!\!\!E_{\rm T}$ system have 
$p_{\rm T} \geq 50 \,{\rm GeV/c}$ has therefore been made.
%

   In addition to the $b$-jet from the SM top decay, we have three
additional jets from the other top decay when the W or 
the chargino decays hadronically.  
We require two jets~\cite{cdf_jetclus} with transverse energy
$E_{\rm T} \ge$ $20 \,{\rm GeV}$ and a third jet with 
$E_{\rm T} \ge$ $15 \, {\rm GeV}$ all within $| \eta | \leq 2.0$.
Jet energy is corrected according to an average response 
function prior to event selection~\cite{cdf_top_kin}.
   We require one of the jets to be identified as a $b$-jet candidate. We 
use a secondary vertex tagging method [1], based on SVX information, 
that reconstructs secondary vertices from $B$ hadron decays.
As in the kinematic top analysis~\cite{cdf_top_kin}, 
we require the three jets to have large polar angles in the rest frame 
of the lepton, $\not\!\!E_{\rm T}$, and jets:
$| \cos ( \theta ^{*} (\mbox{jet}_{i}) ) | < 0.9$,
$| \cos ( \theta ^{*} (\mbox{jet}_{j}) ) | < 0.8$, and
$| \cos ( \theta ^{*} (\mbox{jet}_{k}) ) | < 0.7$, where the jets {\it i,j,k}
are ordered according to $| \cos ( \theta ^{*} )|$.
These requirements were relaxed from the requirement of 
$| \cos ( \theta ^{*} (\mbox{jet}_{i,j,k}) ) | < 0.7$ in~\cite{cdf_top_kin}
in order to obtain a good acceptance for the SM-SUSY top decays.
For simplicity, these polar angles are calculated 
assuming a null longitudinal component for the neutrino.
In order to insure that the three jets are 
well separated in  $\eta$-$\phi$ space,
we require $\Delta R (\mbox{jet, jet}) \geq 0.9$, where 
$\Delta R (\mbox{jet, jet})=\,\sqrt{\Delta \eta^{2} + \Delta \phi ^{2}}$ 
is the minimum distance between the jets.

   With these jet requirements our sample is defined. In the data
$9$ events pass the above cuts.
For the theoretical $t \overline{t}$ production cross section 
of $ 5\, {\rm pb}$~\cite{cdf_cross} we expect $9.5$ events.
Non-top Standard Model background contributes less than  
half an event.

   Jets from a SUSY top decay are less energetic than jets from a SM top decay
due to the presence of LSP's and possible small chargino mass. 
In order to best distinguish between SM and SUSY top decays,
we combine the transverse energy information of the second 
and third most energetic jet in a likelihood variable 
defined as:
$R_{L}= 
\frac{{\mathcal{P}}^{SM-SM}(E_{\rm T}^{jet2})\times {\mathcal{P}}^{SM-SM}(E_{\rm T}^{jet3})}
{ {\mathcal{P}}^{SM-SUSY}(E_{\rm T}^{jet2})\times {\mathcal{P}}^{SM-SUSY}(E_{\rm T}^{jet3})}$, 
where $\mathcal{P}$ are the expected differential transverse 
energy distributions $\frac{1}{\sigma} \frac{d\sigma}{dE_{\rm T}}$ 
evaluated from the Monte Carlo.

\newpage
\begin{center}
\begin{figure}[t!]
\vspace{-0.4cm}
\hspace{-0.7cm}
\begin{minipage}{2.5in}
  \epsfxsize3.6in 
  \hspace*{0.0cm}\epsffile{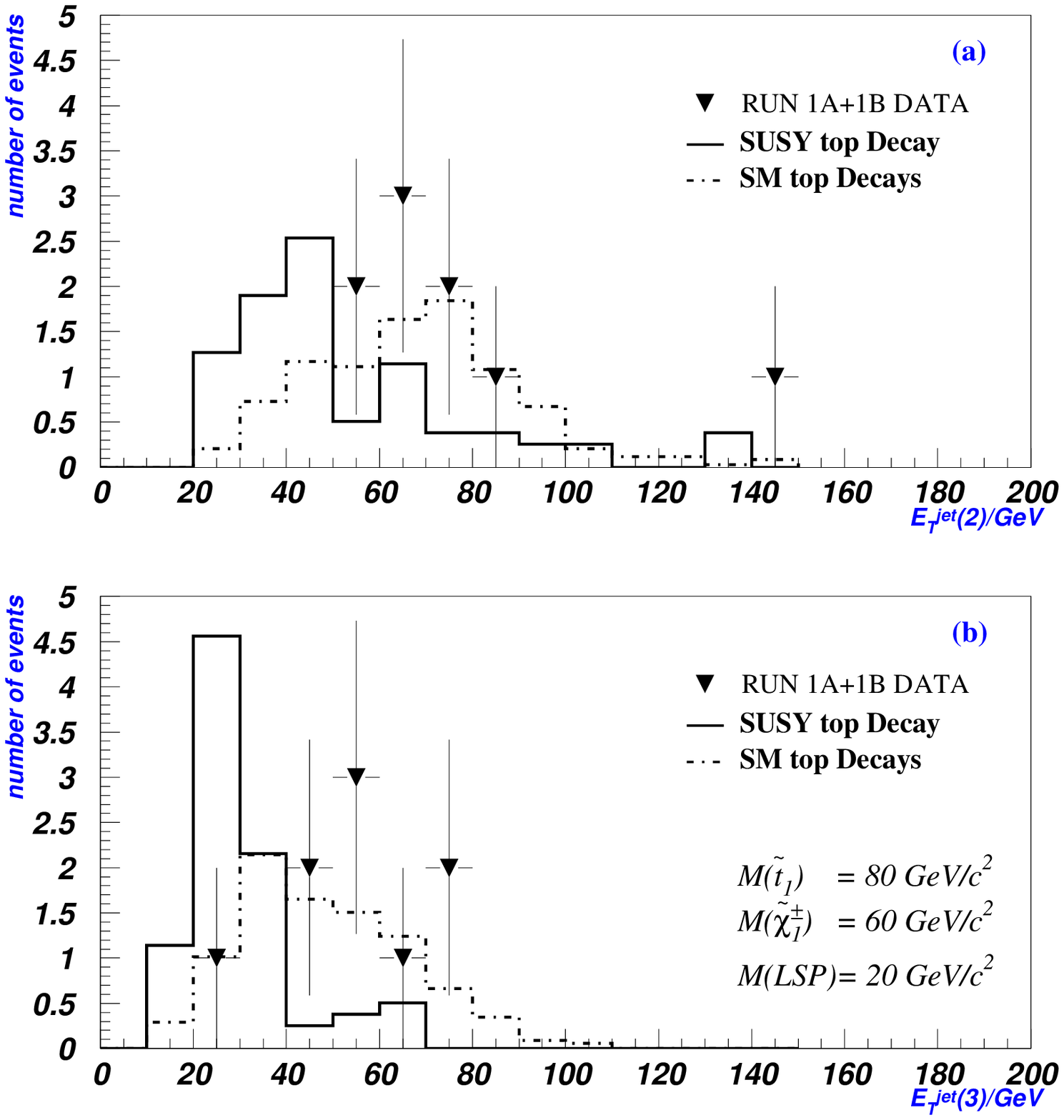}
  \epsfxsize3.6in 
  \hspace*{-0.4cm}\epsffile{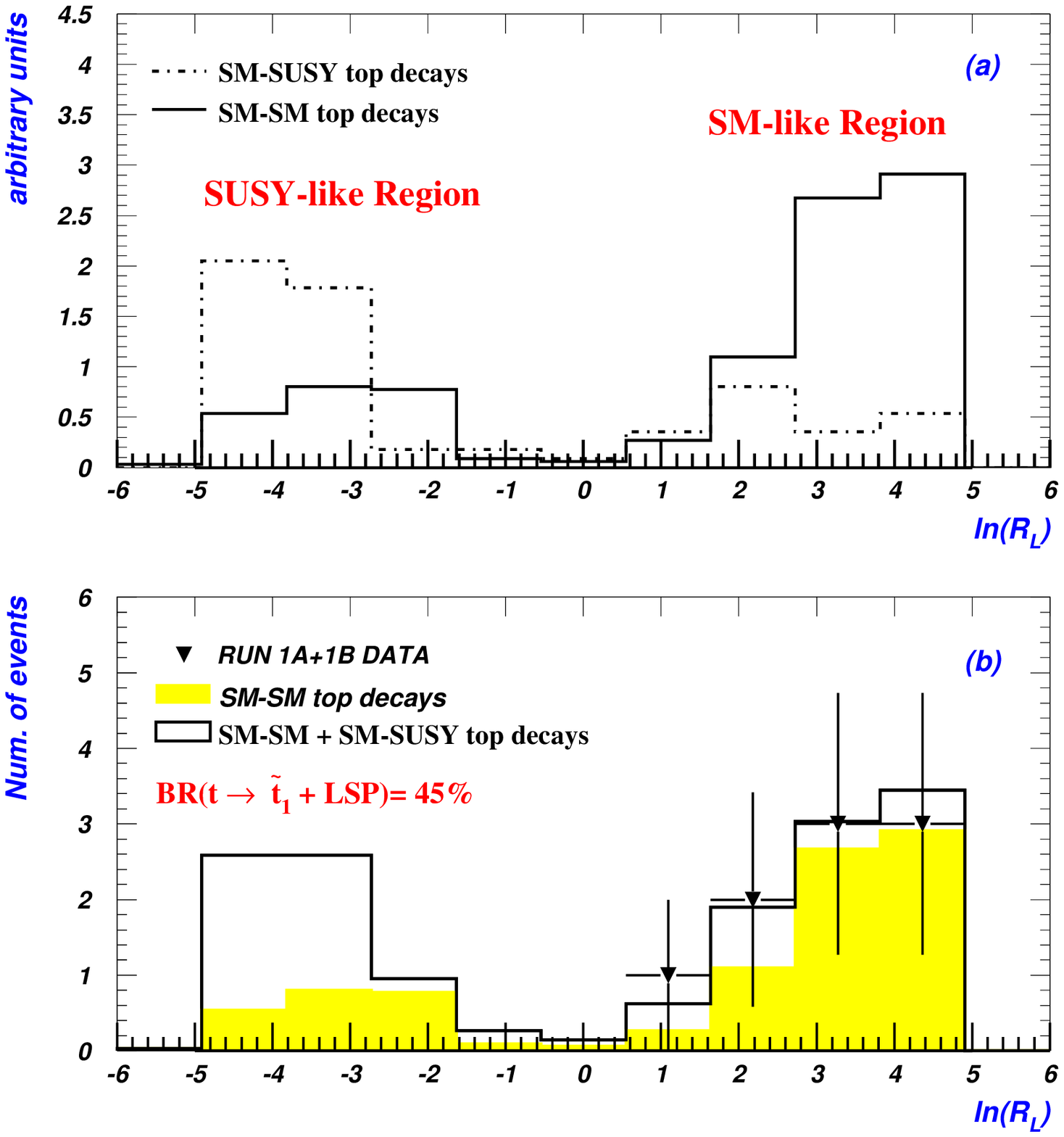}
\end{minipage}\hfill
\end{figure}
\begin{figure}[th!]
\vspace{-0.4cm}
\begin{minipage}{3.1in}
\vspace{-0.4cm}
\caption{\it Transverse Energy distribution of 
a) the second $E^{jet}_{\rm T}(2)$
and b) third jet $E^{jet}_{\rm T}(3)$
for events with two SM top decays and for events with 
one SM and one SUSY top decay.}
\label{f:et23_likelihood}
\end{minipage}\hfill
\hspace{0.5cm}
\begin{minipage}{3.1in}
\vspace{-0.85cm}
  \caption{\it  a) Comparison of ${\rm ln}(R_{L})$ for SM-SM top decays 
and SM-SUSY top decays after all the cuts have been applied.
b) The ${\rm ln}(R_{L})$  distribution for CDF Run 1 data.}
\end{minipage}\hfill
\vspace{-.39cm}
\end{figure}
\end{center}
\vspace{-0.75cm}

\noindent

   Figure 1 shows those $E_{\rm T}$ distributions 
while Figure 2 shows the distribution of the likelihood 
variable for SM and SUSY top decays.
Events with one top quark decaying into stop plus LSP are clustered in the
region of negative values of ${\rm ln}(R_{L})$, 
whereas events with two SM top decays 
are at positive values~\cite{susysusy}.
The region ${\rm ln}(R_{L})<-1$
defines our SUSY search region. The region
${\rm ln}(R_{L})>-1$
is dominated by double SM top decays and will be used to
normalize the expected number of these decays. Our search is then independent
of the $t \overline{t}$ production cross section~\cite{footnote1}.

   To study the distributions of the kinematic variables 
in the supersymmetric top decays, the signals are produced with the
%
{\sc ISAJET} Monte Carlo generator~\cite{isajet} and
passed through detector simulation programs~\cite{QFL}. We have fixed the top quark 
mass to $175 \, {\rm GeV/c}^{2}$ and varied stop, chargino, and LSP masses.
Different branching ratios for 
top decaying into stop plus LSP are obtained by analyzing
the three possible
combinations of top decays (SM-SM, SM-SUSY, and SUSY-SUSY) individually and 
recombining them with appropriate weights.

   Systematic uncertainties in the simulation of $t \overline{t}$ events
can impact both the expected number of events and the shape of\, the\, two 
jet\, $E_{\rm T}$\, distributions\, and\, thus the likelihood\, distribution. 

\newpage
\begin{figure}[t!]
\hspace{-0.8cm}
\begin{minipage}{2.5in}
  \epsfxsize3.6in 
  \hspace*{0.0cm}\epsffile{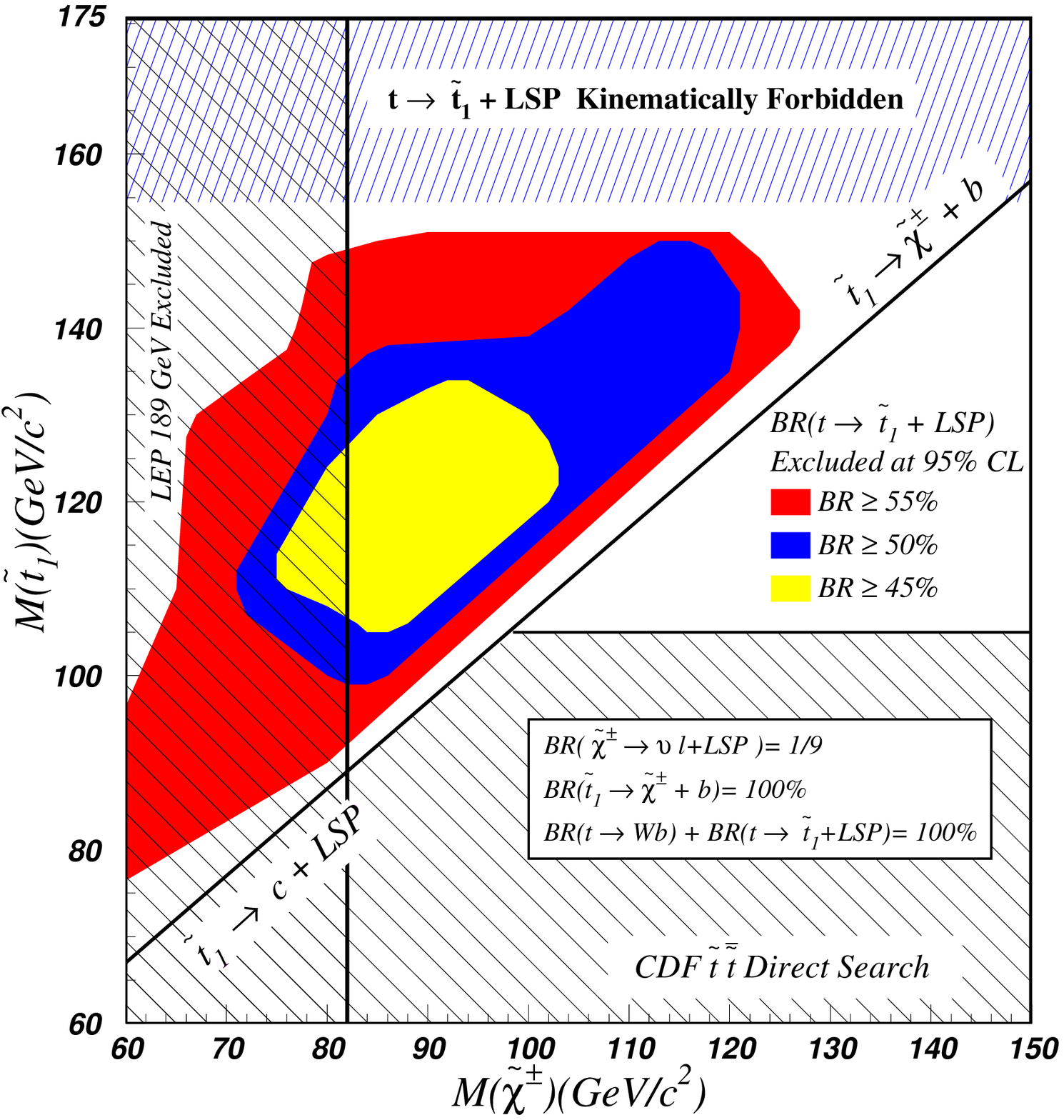}
  \epsfxsize3.6in 
  \hspace*{-0.1cm}\epsffile{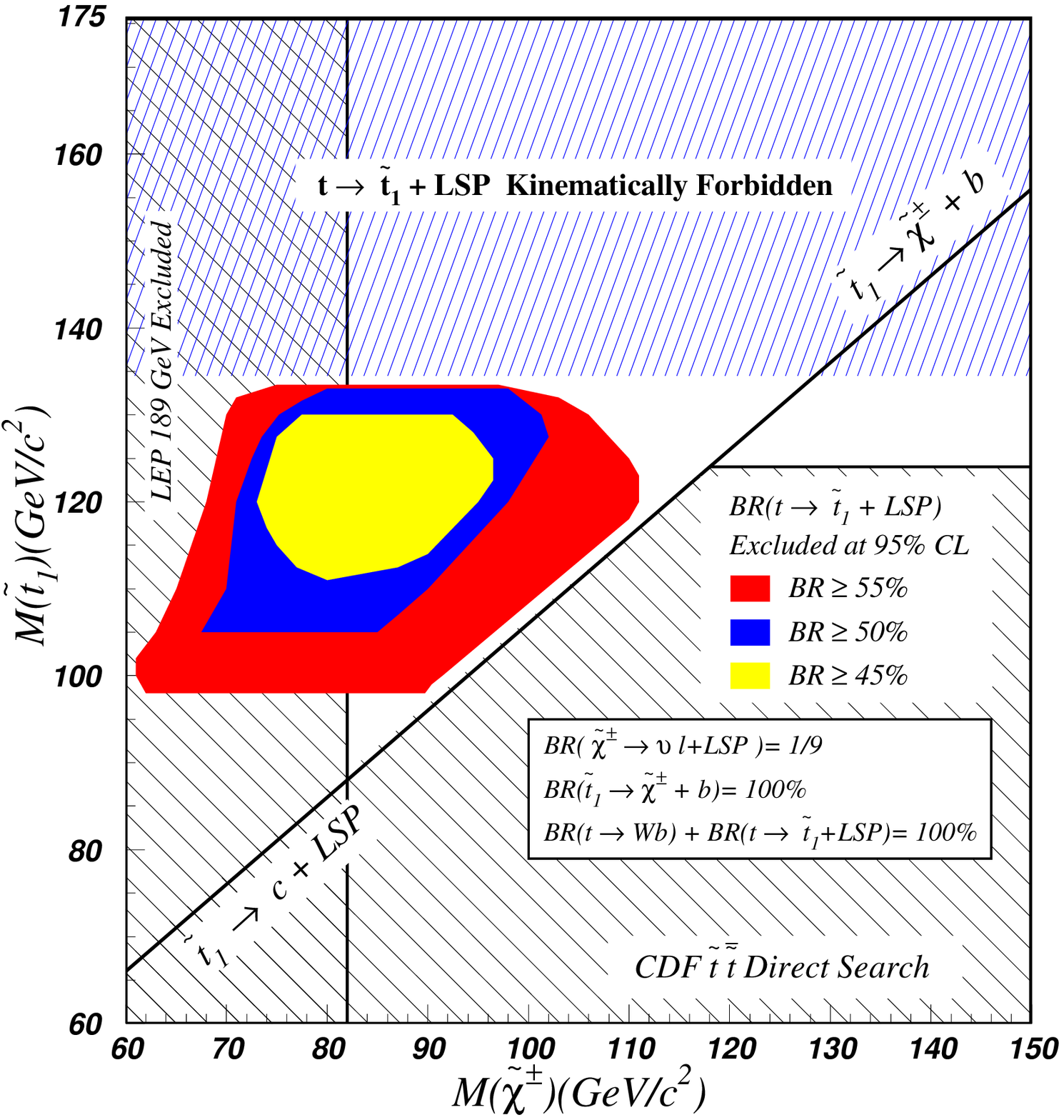}
\end{minipage}\hfill
\caption{\it Top into stop plus LSP branching ratio limits as a function of
stop and chargino masses for a LSP mass of $20 \, {\rm GeV/c}^{2}$ (left) 
and $40 \, {\rm GeV/c}^{2}$ (right).}
\label{f:stop_br_limit}
\vspace{-0.5cm}
\end{figure}
\noindent
We have evaluated the systematic uncertainties for multiple points in 
the parameter space of the stop and chargino masses. 
The systematic uncertainties are expected to become significant 
close to the kinematic bounds.
In the region of small chargino mass the uncertainties due to
gluon radiation and the calorimeter energy scale are important. 
The procedure used to evaluate these systematic uncertainties 
is the same as that of  Ref.~\cite{cdf_top_prd}. The uncertainty due to
the $t \overline{t}$ production cross section and to the integrated luminosity 
is negligible since we normalize the Monte Carlo predictions to the data in the 
SM dominated region of large likelihood.
We note that this normalization has a large statistical 
uncertainty due to the small number of events in this region.
The uncertainty in the mass of the top quark becomes important in the
region of large stop mass, close to the kinematic limit.
The effect of using parton distribution functions 
other than CTEQ-3 (LO)~\cite{cteq} is within the 
statistical uncertainty of the Monte Carlo
samples. The total systematic uncertainty for both expected number of
events and fraction of events with   ${\rm ln}(R_{L})<-1$ 
from SM-SUSY top quark
decays is typically around 40\%.
The major systematic uncertainty 
comes from the calorimeter energy scale and it varies between $\pm$15\%
and $\pm$25\% 
in the region of the analyzed parameter space. The uncertainty
on the top quark mass contributes about  $\pm$15\%  
for most of the parameter space. 
The uncertainty due to the gluon radiation, on both number
of events and shape, is always less than  $\pm$10\%.

   All nine events observed in the data 
cluster in the SM-like region of large   ${\rm ln}(R_{L})$. 
To set a limit on the branching ratio of top decaying into stop plus LSP, 
we calculate the branching ratios that would yield at least one event 
in the SUSY-like region 95\% 
of the time as a function of stop, chargino and LSP mass. 
The method used is essentially a Bayesian-style integration over
the systematic and statistical uncertainties in the SM-SM, SM-SUSY, and
SUSY-SUSY contributions, where the uncertainties are assumed to be 
Gaussian distributed.
   Figure~\ref{f:stop_br_limit} shows the 95\% 
confidence level top decaying into
stop plus LSP branching ratio limit as a function of stop and chargino mass,
for a LSP mass of $20$ and $40 \, {\rm GeV/c}^{2}$. For
larger LSP masses the kinematically allowed region shrinks. The sensitivity
of this analysis, however, stays rather constant.

   In conclusion, we have looked for the top decay 
into stop plus the LSP. 
In the case of a light stop squark, the top is allowed to decay into stop 
plus a LSP. The number of events observed in the data is 
consistent with the Standard Model top decay expectation. 
We exclude branching ratios for top decaying into stop above 45\%
for an LSP mass up to $40$ $ {\rm GeV/c}^{2}$.
   The upcoming run at the Tevatron, which will feature an approximately 
20-fold increase in the total integrated luminosity as well as a 
significantly improved CDF detector, should allow these results to be 
greatly extended.

\vspace{3 ex}

{\noindent \large \bf Acknowledgements}

   We thank the Fermilab staff and the technical staff of the participating
institutions for their vital contributions. This work was supported by the
U.S. Department of Energy and National Science Foundation; the Italian
Istituto Nazionale di Fisica Nucleare; the Ministry of Education, Science
and Culture of Japan; the Natural Sciences and Engineering Research Council
of Canada; the National Science Council of the Republic of China;  and the
A.~P.~Sloan Foundation.


\begin{thebibliography}{99}
\bibitem{cdf_coord}
  In the CDF coordinate system, $\phi$ is the azimuthal angle and 
  $\theta $ is the polar angle with respect
  to the proton beam direction. The pseudorapidity $\eta $ is defined as
  $\eta=- \ln \tan (\theta / 2)$. The transverse momentum of a particle is
  $p_{\rm T} = p \sin \theta $. If the magnitude of this vector is obtained
  using the calorimeter energy rather than the spectrometer momentum, it
  becomes the transverse energy $E_{\rm T}$.
  Jets are defined as clusters of energy in $\eta-\phi $ space with a 
  fixed cone of $0.4$.
  The missing transverse energy ($\not\!\!\!E_{\rm T}$) is defined 
  as the difference between the vector sum of all the transverse 
  energies and zero.
\bibitem{susysusy} 
   Events with two SUSY top decays have small acceptance due to the 
   W requirement in our selection.
\bibitem{footnote1}
  The $t \overline{t}$ production cross section measured by CDF is based on
  the assumption of only SM top decays. Theoretical calculations may omit
  $t \overline{t}$ contributions from processes such as gluino production.
\end{thebibliography}
\end{document}